\documentclass{INTERSPEECH2023}

\usepackage{multirow}
\usepackage{flushend}
\usepackage{makecell}
\usepackage{subcaption}
\usepackage{graphicx}
\usepackage[normalem]{ulem}
\usepackage{microtype}

\interspeechcameraready 

\title{Leveraging Semantic Information for Efficient Self-Supervised Emotion Recognition with Audio-Textual Distilled Models}
\name{Danilo de Oliveira, Navin Raj Prabhu, Timo Gerkmann}
\address{Signal Processing (SP), Universit\"at Hamburg, Germany}
\email{\{danilo.oliveira, navin.raj.prabhu, timo.gerkmann\}@uni-hamburg.de}

\begin{document}
\maketitle
 
\begin{abstract}
In large part due to their implicit semantic modeling, self-supervised learning (SSL) methods have significantly increased the performance of valence recognition in speech emotion recognition (SER) systems. Yet, their large size may often hinder practical implementations.
In this work, we take HuBERT as an example of an SSL model and analyze the relevance of each of its layers for SER.  We show that shallow layers are more important for arousal recognition while deeper layers are more important for valence. This observation motivates the importance of additional textual information for accurate valence recognition, as the distilled framework lacks the depth of its large-scale SSL teacher. Thus, we propose an audio-textual distilled SSL framework that, while having only $\sim$20\% of the trainable parameters of a large SSL model, achieves on par performance across the three emotion dimensions (arousal, valence, dominance) on the MSP-Podcast v1.10 dataset.

\end{abstract}

\section{Introduction}

Speech signals carry rich information on an individual's emotional states, expressed through both paralinguistic and semantic cues. Backed by the circumplex model \cite{russell1980circumplex}, research on speech emotion recognition (SER) is typically performed by studying the emotional expressions across multiple dimensions, namely the activation-deactivation dimension (\emph{arousal}), the pleasure-displeasure dimension (\emph{valence}), and the speaker confidence-diffidence dimension (\emph{dominance}). Prior works in literature reveal that certain speech cues carry more information on a particular emotional dimension. For instance, the arousal dimension is well explained by paralinguistic cues \cite{Prabhu2021EndToEndLU}, and semantic cues are more informative of the valence dimension \cite{tzirakis2021-semspeech}. This calls for techniques that learn both paralinguistic and semantic cues simultaneously.

The self-supervised learning (SSL) paradigm is a good candidate to fulfill these requirements. Methods from this domain leverage unlabeled data in a pre-training stage as a way of learning robust {representations} of an input's underlying structure. The pre-training of SSL models is usually succeeded by a fine-tuning stage, in which the model is fed labeled data and trained in a supervised manner for a target downstream task. Following the success of BERT \cite{devlin_bert_2019} in the natural language processing (NLP) domain, models like wav2vec 2.0 \cite{baevski_wav2vec_2020} and HuBERT \cite{hsu_hubert_2021} have provided great performance boosts in speech tasks, ranging from automatic speech recognition (ASR) to speaker identification (SI) \cite{vaessen_fine-tuning_2022} and SER \cite{pepino_emotion_2021, wagner_dawn_2022}.

While SER systems typically perform well in terms of arousal and dominance, modeling valence from speech signals is a challenging task \cite{kusha2022_extValence, Prabhu2021EndToEndLU}, resulting in a performance gap between arousal/dominance and valence estimation. Prior works have successfully managed to reduce this gap by fine-tuning wav2vec 2.0 and HuBERT for emotion recognition \cite{wagner_dawn_2022}. The improved valence prediction results come from the fact that these transformer-based SSL models can \textit{implicitly} capture semantic content present in speech, along with the paralinguistics \cite{triantafyllopoulos_probing_2022}. However, such models still underperform on valence compared with models that \textit{explicitly} include semantic information through BERT-encoded features \cite{srinivasan_representation_2022}. These findings suggest that the semantic information is not completely modeled by speech-only SSL techniques.

Despite the reduced performance gap between arousal and valence, an important drawback of SSL models is their size: the smallest versions of BERT, wav2vec 2.0, and HuBERT contain 110, 95, and 90 million parameters, respectively, making their use costly or even prohibitive in some cases. Knowledge distillation \cite{hinton_distilling_2015} methods have found success in addressing this issue, in particular in the NLP domain \cite{sanh_distilbert_2020, jiao_tinybert_2020}. The goal of these methods is to compress the knowledge acquired by a large network (the teacher) by transferring it to a smaller one (the student). Applied to HuBERT, this framework results in DistilHuBERT \cite{chang_distilhubert_2022}, with only 25\% of the parameters of its teacher.

An analysis of DistilHuBERT suggests that the distilled network is good at encoding paralinguistic information, given the high importance of its representations in the SI task \cite{chang_distilhubert_2022}; this indicates potentially good performance in arousal estimation. Moreover, the inner representations of wav2vec 2.0 have an acoustic-linguistic hierarchy \cite{pasad_layer-wise_2021}, where the information embedded in each layer output mutates with network depth, from an acoustic to a semantic nature. These findings further motivate us to investigate how well the distilled model fares in predicting each of the emotional dimensions, especially valence. 

In this paper, we make the following contributions: We show that the shallow layers of SSL models are more important for arousal recognition while the deeper layers are more important for valence recognition. We argue that the students in distilled models like DistilHuBERT lack the required depth to properly model valence. Therefore, we show that adding textual information is particularly helpful for distilled SSL networks like DistilHuBERT, considerably more important than for non-distilled large-scale SSL networks. To the best of our knowledge, we are the first in the literature to use distilled SSL models in arousal, valence and dominance modeling and to analyze the implication of the distillation process (i.e., layer selection and compression) towards bridging the gap between valence and arousal estimation.

\section{Proposed methodology}\label{sec:methodology}

The task of emotion recognition is formulated here as follows: given a single-channel audio input containing a spoken utterance $\mathbf{X}_A \in\mathbb{R}^{1\times S}$, where $S$ is the number of samples, we want to estimate three emotional expression scalar values: arousal ($Y_a$), valence ($Y_v$) and dominance ($Y_d$). Our SER model $f_\mathrm{A}$ should map the input utterance $\mathbf{X}_A$ to an estimate of the three emotion dimensions, simultaneously: 
\begin{equation}\label{eq:audio_model_eq}
\widehat{\mathbf{Y}} = f_{\mathrm{A}}(\mathbf{X}_A),
\end{equation}
where $\widehat{\mathbf{Y}} = \mathrm{concat}(\widehat{Y}_a, \widehat{Y}_v, \widehat{Y}_d)$.

In the multi-modal case, we also have text as an additional input. The tokenized text is denoted by $\mathbf{X}_T \in\mathbb{N}^{1\times N}$, where $N$ is the number of tokens in the utterance. The audio-textual model mapping is therefore
\begin{equation}\label{eq:audio_text_model_eq}
\widehat{\mathbf{Y}} = f_\mathrm{A+T}(\mathbf{X}_A, \mathbf{X}_T).
\end{equation}

Figure~\ref{fig:system} illustrates the models. They contain essentially audio and text encoders ($\mathrm{SSL}_A$ and $\mathrm{SSL}_T$, respectively), pooling operations and a feed-forward regression head (FFN). The pooling block consists of average- and max-pooling across the sequence dimension. The FFN layer contains two fully-connected layers and a hyperbolic tangent activation function. Its final layer has three output features to estimate $\widehat{\mathbf{Y}}$. 

\subsection{Audio-only SSL framework}

The audio-only system is depicted in Figure~\ref{fig:audio_model}. Aiming at extracting features from the time-domain input $\mathbf{X}_A$, we employ pre-trained models as $\mathrm{SSL}_A$, namely wav2vec 2.0, HuBERT and DistilHuBERT. They share the same structure: a convolutional encoder followed by transformer blocks. The convolutional encoder creates frames of latent representations that act as tokens for the creation of contextualized features $\mathbf{C}_A\in\mathbb{R}^{d_A\times K}$ by the transformer encoder, where $d_A$ is the hidden dimension size and $K$ is the number of frames, dependent on the configuration of the convolutional blocks. The transformer encoder is similar to its text counterpart, presented in the next section. 

Wav2vec 2.0 uses product quantization to generate a finite set of representations and compares them against the context representations using a contrastive loss. HuBERT, on the other hand, is trained in two separate steps: the first is an offline step that creates discrete pseudo-labels by clustering audio-based features, and the second consists in masked prediction of cluster assignments. Finally, DistilHuBERT is a model obtained by distilling HuBERT through a framework of three prediction heads that aim at estimating HuBERT's 4th, 8th and 12th layers. Though the models make use of largely different pre-training techniques, during fine-tuning/inference their architectures differ essentially in the number of layers and hidden size.

In order to use the extracted representations for our emotion recognition task, the pooling block compresses them to an utterance-level representation $\mathbf{P}_A\in\mathbb{R}^{d_A\times 1}$. This is then fed to the FFN head, which outputs the estimates for arousal, valence and dominance.

\subsection{Audio-textual SSL framework}\label{subsec:text_ssl}

For experiments including text inputs, the token IDs from the tokenized text $\mathbf{X}_T$ pass through an embedding layer, and are then fed to a BERT-like encoder. This class of language models is pre-trained by randomly masking tokens and attempting to predict them using the unmasked ones as context. The architecture is essentially a sequence of bidirectional transformer layers. The resulting vector $\mathbf{C}_T\in\mathbb{R}^{d_T\times N}$, with $d_T$ being the size of the text hidden dimension, is an embedding enriched by contextual information.

In our experiments, we use TinyBERT \cite{jiao_tinybert_2020} as $\mathrm{SSL}_T$, with the intent of minimizing overhead. It is a distilled version of BERT, trained by applying distillation in two steps: pre-training and fine-tuning, aiming at reproducing BERT's capabilities of generalization and downstream task performance, respectively. The authors report 96.8\% of the performance of BERT base on an NLP task benchmark, even though the model is 7.5x smaller.

\begin{figure}
    \begin{subfigure}{0.5\textwidth}
        \centering
        \includegraphics[scale=0.9]{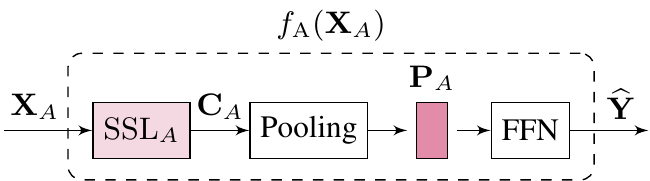}
        \subcaption{Audio-only model}
        \label{fig:audio_model}
    \end{subfigure}
    \par\medskip
    \begin{subfigure}{0.5\textwidth}
        \centering
        \includegraphics[scale=0.9]{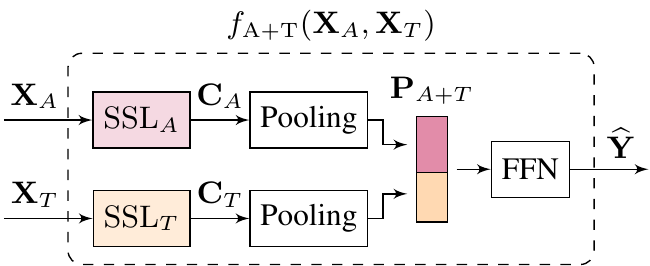}
        \subcaption{Audio + text model}
        \label{fig:audio_text_model}
    \end{subfigure}
    \caption{System architecture, mapping a spoken utterance $\mathbf{X}_A$ (and, in the audio-textual case, its corresponding transcript $\mathbf{X}_T$) to a set of arousal, valence and dominance estimates $\widehat{\mathbf{Y}}$.}
    \label{fig:system}
\end{figure}

Inspired by \cite{pepino_fusion_2020, siriwardhana_jointly_2020, wagner_dawn_2022}, we use a simple concatenation method for the fusion of modalities, shown in Figure~\ref{fig:audio_text_model}: the pooled features from each encoder are concatenated in the feature dimension, resulting in $\mathbf{P}_{A+T}\in\mathbb{R}^{(d_A+d_T)\times 1}$. We then pass it through a feed-forward block similar to the audio-only case, only with the number of input features adjusted to incorporate the additional text representations.

\section{Implementation details}

\begin{table*}[hbt!]
  \caption{CCC scores on MSP Podcast (v1.10). Seen scenarios is the test set containing segments from podcasts included in the train or development sets, while the unseen scenarios set contains segments from podcasts not present in any other partition.}
  \label{tab:ccc_a_t}
  \centering
  \begin{tabular}{c|c c|c c c|c c c}
    \toprule
    \multirow{2}{*}{\textbf{Modality}} & 
    \multirow{2}{*}{\textbf{Model}} & 
    \multirow{2}{*}{\textbf{\makecell{\#Params\\(M)}}} & 
    \multicolumn{3}{c|}{\textbf{Seen Scenarios}} & 
    \multicolumn{3}{c}{\textbf{Unseen Scenarios}} \\
    & & &
    \textbf{Arousal} &
    \textbf{Valence} & 
    \textbf{Dominance} &
    \textbf{Arousal} &
    \textbf{Valence} & 
    \textbf{Dominance} \\
    \midrule
    \multirow{3}{*}{Audio}        & w2v2-L-robust(p) & 165         & \textbf{0.627} & \textbf{0.470} & \textbf{0.521} & 0.462          & \textbf{0.263} & 0.390          \\
                                  & HuBERT base      & 95          & 0.608          & 0.440          & 0.486          & 0.388          & 0.225          & 0.344          \\
                                  & DistilHuBERT     & \textbf{24} & 0.622          & 0.328          & 0.513          & \textbf{0.470} & 0.169          & \textbf{0.395} \\\hline
    \multirow{3}{*}{Audio + Text} & w2v2-L-robust(p) & 180         & \textbf{0.619} & \textbf{0.560} & 0.484          & 0.469          & \textbf{0.347} & 0.353          \\
                                  & HuBERT base      & 109         & 0.613          & 0.532          & 0.493          & 0.461          & 0.315          & \textbf{0.396} \\
                                  & DistilHuBERT     & \textbf{39} & 0.614          & 0.519          & \textbf{0.509} & \textbf{0.475} & 0.333          & 0.392          \\      
    \bottomrule
  \end{tabular}
\end{table*}

\subsection{Dataset}

For training, validation and testing, we use the MSP-Podcast dataset \cite{lotfian_building_2019}, version 1.10. It contains approximately 166 hours of audio extracted from podcast data, labeled at utterance level for arousal, valence and dominance on a scale of 1 to 7. V1.10 features human-labeled transcripts, which serve as our oracle text information. The dataset is split into four partitions: \texttt{train}, \texttt{development}, \texttt{test1} and \texttt{test2}. Differently from \texttt{test1}, the segments in \texttt{test2} originate from podcasts not present in any other partitions. We therefore consider it our \textit{unseen scenarios} set, while \texttt{test1} is our \textit{seen scenarios} test data.

\subsection{Baseline and proposed models}

In our experiments, we propose a framework based exclusively on distilled models and compare it against \textit{base} and \textit{large} $\mathrm{SSL}_A$ baselines. As the \emph{base} $\mathrm{SSL}_A$ model, we use HuBERT base, which has 12 transformer layers and hidden size $d_A = 768$ \cite{hsu_hubert_2021}. As the \emph{large} $\mathrm{SSL}_A$ model, we use the \textit{pruned} w2v2-L-robust from \cite{wagner_dawn_2022}, which we denote as w2v2-L-robust(p). It uses the first 12 (out of 24) layers of the original model, with a hidden size $d_A = 1024$. As the proposed distilled $\mathrm{SSL}_A$, we use the DistilHuBERT model, that has the same hidden size as HuBERT ($d_A = 768$) but contains only 2 layers. For our audio-textual experiments, as the distilled $\mathrm{SSL}_T$, we employ the 4-layer TinyBERT encoder, with $d_T = 312$. 

We make use of the pre-trained SSL models available on Huggingface\footnote{https://huggingface.co}. When fine-tuning, we follow the usual procedure of freezing the weights of the convolutional encoders present in the speech models. Emotional expression target labels are normalized to $[0,1]$. We use Adam \cite{kingma_adam_2015} as the optimizer, with a learning rate of $10^{-5}$ and early stopping to prevent overfitting. We employ dropout of $0.1$ throughout the model. The training data are batched with batch size 16, using buckets sorted by length in order to reduce the amount of padding. The networks are fine-tuned on a single NVIDIA A6000 GPU.

\subsection{Loss function}
To train the proposed architecture, we use the concordance correlation coefficient (CCC) loss \cite{lin_concordance_1989}, a widely used loss function in SER research \cite{schuller_speech_2018, Prabhu2021EndToEndLU, tzirakis2021-semspeech}. The CCC measures the similarity between two variables and varies between $-$1 and $+$1, where $+$1 denotes perfect similarity and 0 denotes perfect orthogonality between the variables. For Pearson's correlation coefficient $\rho$, the CCC between $\mathbf{y}$ and its estimate $\widehat{\mathbf{y}}$ is formulated as
\begin{equation}\label{loss:CCC}
    \mathcal{L}_{\text{CCC}}(\mathbf{y}, \widehat{\mathbf{y}}) = {\frac {2\rho \sigma_{\mathbf{y}}\sigma_{\widehat{\mathbf{y}}}}{\sigma _{\mathbf{y}}^{2}+\sigma_{\widehat{\mathbf{y}}}^{2}+(\mu _{\mathbf{y}}-\mu _{\widehat{\mathbf{y}}})^{2}}},
\end{equation}
where $\mu$, and $\sigma$ are the mean and standard deviation, respectively, of the corresponding variables. From \eqref{loss:CCC}, it can be noted that the CCC takes both the linear correlation and the bias between $\mathbf{y}$ and $\widehat{\mathbf{y}}$ into consideration while quantifying their similarity, hence it is preferred over the Pearson correlation as a loss function for SER. Our training minimizes $1 - \mathcal{L}_{\text{CCC}}(\mathbf{y}, \widehat{\mathbf{y}})$ averaged across the three emotional dimensions. 

\section{Results and Discussion}

\subsection{Explicit inclusion of semantic information}

We evaluate our audio-only and audio-textual versions of DistilHuBERT along with the HuBERT base and w2v2-L-robust(p) baselines. The results are shown in Table~\ref{tab:ccc_a_t}. Firstly, from the \textit{audio-only} models, it can be observed that DistilHuBERT manages to match the large SSL baseline w2v2-L-robust(p), for both seen and unseen scenario test data even though it has only $\sim$15\% of the total number of parameters. However, the lack of modeling depth and capacity seems to heavily impact valence estimation. The HuBERT base version lags behind in arousal and dominance, but has valence scores closer to those of w2v2-L-robust(p).

Secondly, the \textit{audio-textual} models considerably outperform their audio-only counterparts in terms of valence estimation, even in the case of w2v2-L-robust(p), whose dominance modeling performance decreased, while arousal remained at roughly the same level. The audio-textual versions of the base and distilled HuBERT models either improved or remained at the same level in all cases. Notably, the audio-textual distilled model performs on par with the larger model not only in arousal and dominance but also valence, thus confirming the effectiveness of the explicit inclusion of text, particularly for distilled models.

\subsection{Layer importance for the emotional dimensions}

\begin{figure}
\includegraphics{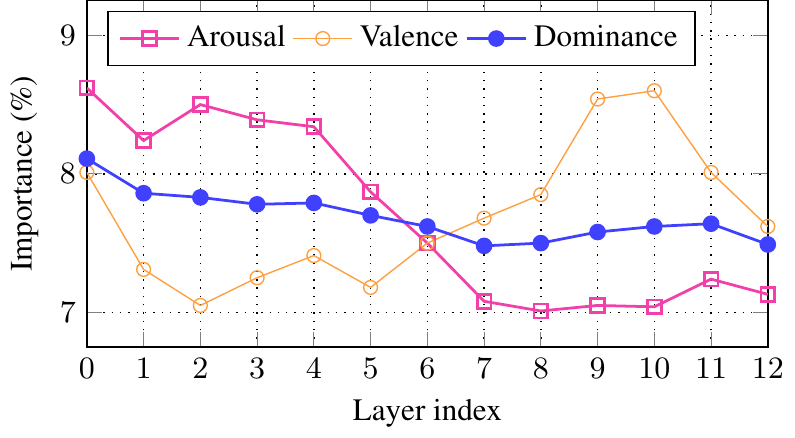}
\caption{Normalized importance given to each layer of HuBERT when training in a frozen weights setting. Index 0 corresponds to the output of the convolutional encoder. The subsequent indexes reference each of the transformer encoder's 12 layers.}
\label{fig:plot_layers}
\end{figure}

As presented in Table~\ref{tab:ccc_a_t}, the audio-only DistilHuBERT, despite its performance in arousal estimation, performs poorly in terms of valence. To explain this behavior, and also to better understand what kind of information the distilled model learns from the teacher, we proceed by analyzing the relevance of the transformer layers of the pre-trained HuBERT base encoder for arousal, valence and dominance modeling, this time individually. In a similar fashion to \cite{pepino_emotion_2021, chang_distilhubert_2022}, we perform a weighted sum of the outputs of the encoder's layers. The weights are normalized to sum to one, and the encoder weights are kept frozen while training the regression head for estimating the target emotional expression. The resulting layer-weighting parameter values found during training are plotted in Figure~\ref{fig:plot_layers}. Index 0 corresponds to the convolutional encoder's outputs, and 1-12 are each of the transformer layers.

Figure~\ref{fig:plot_layers} reveals that for arousal estimation the first layers up to layer 4 are more important than deeper layers, which hints at the specialization of the first layers on paralinguistic features. This coincides with the findings of \cite{pasad_layer-wise_2021} for the similar wav2vec 2.0 model. It is interesting to see that the layer importance follows an opposite trend for valence: Here, lower importance is observed in the first few transformer blocks while deeper layers are more important to model valence. A peak is observed at layers 9 and 10. It raises a concern for the use of DistilHuBERT, since it lacks depth to model the deeper layers.
Additionally, its two layers are initialized from the first two of HuBERT. This can explain the distilled model's performance gap between arousal and valence estimations. Furthermore, it also motivates the explicit inclusion of text in SSL models, especially for the distilled models.

\subsection{Fine-tuning vs. freezing}

\begin{figure}[t!]
    \includegraphics[scale=0.9]{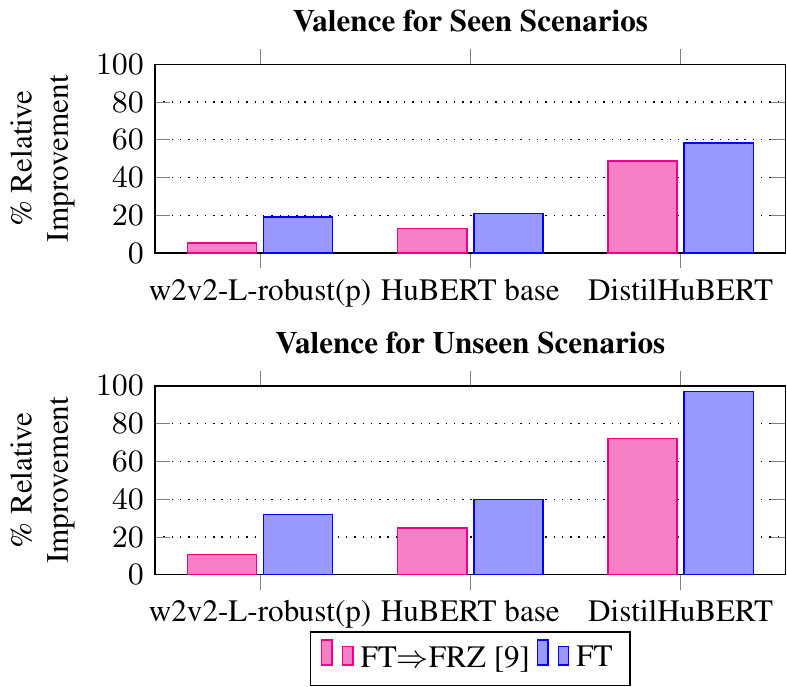}
    \caption{Relative valence CCC improvement of audio+text models  over the audio-only case. ``FT$\Rightarrow$FRZ'' corresponds to the method used in \cite{wagner_dawn_2022}. ``FT'' refers to the fine-tuning used in this paper.}
    \label{fig:valence_boost}
\end{figure}

Motivated by related works in audio-textual emotion recognition that either fine-tune the SSL encoders' weights \cite{siriwardhana_jointly_2020} or keep them frozen while only training a regression head \cite{pepino_fusion_2020, wagner_dawn_2022, srinivasan_representation_2022}, we run experiments to compare how much each approach helps to improve valence predictions. Figure~\ref{fig:valence_boost} displays the relative improvement of the valence CCC score over the audio-only case for each of the considered models: large-pruned, base and distilled. ``FT'' indicates a model with fine-tuned SSL encoders, while ``FT$\Rightarrow$FRZ'' represents the fusion process used in \cite{wagner_dawn_2022}: a pre-trained speech model is fine-tuned for emotion recognition, then its encoder weights are frozen and used alongside an also frozen text encoder to train a regression head. 

We observe in Figure~\ref{fig:valence_boost} that textual information is particularly helpful for smaller models.
This indicates that larger capacity and/or more training data help the model learn some level of semantic information even without explicit textual input. Nevertheless, in all cases the performance is increased by including textual information, confirming the fact that there's still relevant information in the textual input that is not captured by pre-training with audio alone. Furthermore, we observe that fine-tuning brings even larger improvements in all cases; in particular, for the case of the distilled model, valence estimation performance was double that of the audio-only setting in the unseen scenario test set. We therefore hypothesize that fine-tuning the text model helps it focus on modeling the specific semantic information missing from the speech representation.

\subsection{Quality of transcripts}

When implementing systems that make additional use of text, the audio needs to be transcribed, most practically via an automatic speech recognition (ASR) system. Although practical, ASR models may introduce transcription errors. In order to examine the robustness of the text-informed framework, we run our proposed distilled-only, fine-tuned framework using text transcribed by two ASR systems with different model sizes, and compare it with the model trained on human transcriptions. 

The results are presented in Table~\ref{tab:asr_quality}. We can see that the performance is maintained across ASR models, irrespective of their number of parameters. This suggests that the framework is robust with respect to text, as long as the main semantic information is preserved. This claim is further backed by the fact that performance of ASR-generated transcripts is as good as that of human transcripts. It should be noted that the models are run on clean speech; noisy and reverberant conditions are expected to worsen the quality of transcripts and should be addressed by future studies.

\begingroup
\setlength{\tabcolsep}{4pt}
\begin{table}
  \caption{CCC performance of the fine-tuned DistilHuBERT + TinyBERT system with different transcription methods. 
 ``A'', ``V'' and ``D'' denote arousal, valence and dominance, respectively.}
  \label{tab:asr_quality}
  \centering
  \begin{tabular}{c@{  }c|c|c c c}
    \toprule
    \multirow{2}{*}{\textbf{\makecell{Transcription\\method}}} & \multirow{2}{*}{\textbf{\makecell{\#Params\\(only ASR)}}} & \multicolumn{4}{c}{\textbf{Seen Scenarios}}\\
    & & \textbf{WER} & \textbf{A} & \textbf{V} & \textbf{D} \\
    \midrule
    Human                                   & $-$ &   $-$     & 0.614 & 0.519 & 0.509 \\
    Whisper base \cite{radford_robust_2022} & 74M  & 22.2\% & 0.620 & 0.521 & 0.511 \\
    Whisper tiny \cite{radford_robust_2022} & 39M  & 24.1\% & 0.618 & 0.524 & 0.510 \\
    \bottomrule
  \end{tabular}
\end{table}
\endgroup

\section{Conclusions}

In this study we proposed an audio-textual emotion recognition framework based on distilled models. We highlighted the particular importance of multi-modal audio and text inputs for robust arousal, valence and dominance estimation when using our distilled model.
Despite having only $\sim$20\% of the trainable parameters of the largest baseline, the proposed framework's performance is on par with base and large models not only on seen scenarios, but importantly also on unseen scenario data. We investigated the relevance of HuBERT's inner representations to each of the three emotion dimensions and found the initial layers to be more important for arousal modeling, while the deeper layers focus on information instrumental to valence estimation. This analysis further validates the need for text as extra input to distilled networks for improved valence modeling, as these shallow models cannot extract semantic information from speech as easily as their teacher counterparts. 
Lastly, we confirmed the robustness of our audio-textual network by training it on machine-transcribed audio-text pairs, without loss of performance. Distillation is a promising way to make SSL models more practical, but it is necessary to ensure that both paralinguistic and semantic information is available in order to have robust arousal and valence estimation.

\clearpage

\bibliographystyle{IEEEtran}
\bibliography{references}

\end{document}